# Introduction to Drone Detection Radar with Emphasis on Automatic Target Recognition (ATR) technology

Jiangkun Gong, Jun Yan*, Deyong Kong, and Deren Li

*Abstract*— This paper discusses the challenges of detecting and categorizing small drones with radar automatic target recognition (ATR) technology. The authors suggest integrating ATR capabilities into drone detection radar systems to improve performance and manage emerging threats. The study focuses primarily on drones in Group 1 and 2. The paper highlights the need to consider kinetic features and signal signatures, such as micro-Doppler, in ATR techniques to efficiently recognize small drones. The authors also present a comprehensive drone detection radar system design that balances detection and tracking requirements, incorporating parameter adjustment based on scattering region theory. They offer an example of a performance improvement achieved using feedback and situational awareness mechanisms with the integrated ATR capabilities. Furthermore, the paper examines challenges related to one-way attack drones and explores the potential of cognitive radar as a solution. The integration of ATR capabilities transforms a 3D radar system into a 4D radar system, resulting in improved drone detection performance. These advancements are useful in military, civilian, and commercial applications, and ongoing research and development efforts are essential to keep radar systems effective and ready to detect, track, and respond to emerging threats.

## I. INTRODUCTION

Small drones possess distinctive characteristics, including a low radar cross-section (RCS), slow speeds, and low altitudes [1]. These drones generally fall under Group 1 &2 (refer to Table 1), as designated by the U.S. Department of Defense, which mandates the use of rotating blades for aerial flight [2]. Group 1 & 2 drones typically exhibit an RCS ranging from 0.01 to 0.1 $m^2$, making them approximately 1/10,000 to 1/1,000 the size of a typical airplane. In 2008, the concept of Low, Small, Slow (LSS) radar targets was introduced to describe small airborne objects with a general RCS value below 2 $m^2$, flying at speeds below 200 km/h, and operating at altitudes below 1000 m. This classification is based on the drones' physical attributes, particularly their propulsion systems. Group 1 & 2 drones are commonly lightweight, compact, and employed for short-range reconnaissance and surveillance missions. Furthermore, they find extensive use in civilian applications such as aerial photography, mapping, and inspection. Operated remotely, these drones can be equipped with diverse sensors, cameras, and payloads to gather data and accomplish specific tasks. Overall, Group 1 & 2 drones assume a vital role in contemporary military and civilian operations, offering a cost-effective and versatile platform for a wide array of applications.

Table 1. Drone classification according to the US Department of Defense (DoD)[1,2].

| Category | Maximum Gross Takeoff Weight (Pounds) | Normal Operating Altitude (ft) | Airspeed (Knots) |
|---|---|---|---|
| Group 1 | < 20 | < 1200 AGL[3] | <100 |
| Group 2 | 21–55 | < 3500 AGL | <250 |
| Group 3 | <1320 | <18,000 MSL[4] | <250 |
| Group 4 | >1320 | <18,000 MSL | Any airspeed |
| Group 5 | >1320 | >18,000 MSL | Any airspeed |

1. Source: "Eyes of the Army", U.S. Army Roadmap for UAS 2010–2035 [2]. https://home.army.mil/rucker/index.php, accessed on 8, May, 2023.
2. If the drone has even one characteristic of the next level, it is classified in that level.
3. AGL = Above Ground Level.
4. MSL = Mean Sea Level.

The proliferation of drone threats in both civil and military applications has become a significant concern, evident in various incidents. One such incident occurred in 2018 and 2019 when drones infiltrated Gatwick Airport in London, UK, resulting in severe flight disruptions [3]. The Gatwick drone incident served as a wake-up call for airports and aviation authorities worldwide, highlighting the critical need for robust security measures and strategies to counter potential drone threats. Drones also have also played significant roles in conflicts such as the Nagorno-Karabakh conflict in 2021 and the Ukraine wars in 2022 [4][5][6]. Consequently, the development of Counter Unmanned Aerial System (C-UAS) solutions has rapidly gained momentum in many influential nations.

China launched the "Invisible Sword" anti-unmanned aircraft challenge in 2018, which has been conducted annually for several years. The organizer has rigorously tested the

Jiangkun Gong, Jun Yan, and Deren Li are with State Key State Key Laboratory of Information Engineering in Surveying, Mapping and Remote Sensing, Wuhan University, No. 129 Luoyu Road, Wuhan, China (e-mail: gjk@whu.edu.cn, yanjun_pla@whu.edu.cn, drli@whu.edu.cn. Deyong Kong is with School of Information and Communication Engineering, Hubei Economic University, No.8 Yangqiaohu Road, Wuhan, China (e-mail: kdykong@hbue.edu.cn)

*Corresponding author: Jun Yan (yanjun_pla@whu.edu.cn; +86-027-68778527)

solutions provided by over 100 domestic anti-unmanned aircraft manufacturers involved in technology and product development. Through these assessments, they have gained a comprehensive understanding of the technical landscape in the domestic anti-unmanned aircraft technology field. These efforts have helped debunk false claims made by certain manufacturers, rectify the industry atmosphere, and lay a solid foundation for the advancement of anti-unmanned aircraft technology. The U.S. established the Joint Counter-Small Unmanned Aircraft Systems Office (JCO) with the U.S. Army taking the lead in December 2019. The primary responsibility of the JCO is to oversee all anti-unmanned aircraft development projects within the U.S. military. It works in collaboration with various combatant commands and the Office of the Deputy Secretary of Defense, responsible for procurement, to conduct testing and evaluation of deployed anti-unmanned aircraft projects. These evaluations help determine the future development direction and standards for anti-unmanned aircraft projects within the U.S. military. The establishment of this joint leadership organization has provided a conducive environment for the rapid advancement of the U.S. military's anti-unmanned aircraft system. According to the latest report on "Department of Defense Counter-Unmanned Aircraft Systems" by the Congressional Research Service of the United States [7], published on May 31, 2022, the U.S. Department of Defense has planned to invest a minimum of 668 million U.S. dollars in research and development of counter-unmanned aircraft systems (C-UAS) technology for the fiscal year 2023. Additionally, the budget allocated for the procurement of anti-unmanned aircraft weapons is expected to exceed 78 million U.S. dollars.

In the meantime, numerous commercial C-UAS solutions are currently available in the market. Among these solutions, two prominent examples are the AUDS (Anti-UAV Defense System) and the Drone Dome.

The AUDS system, developed in 2015 by a consortium of UK defense companies, exemplifies a comprehensive C-UAS solution. It integrates various components, including radar, electro-optic sensors, and directional RF suppression/interference systems. Specifically, Blighter's A400 series Ku-band electronic scanning radar employs a modular design incorporating a high-efficiency passive electronic scanning array (PESA) and frequency-modulated continuous wave (FMCW) technology. The system's RF suppression capabilities selectively or simultaneously activate within the 400MHz to 6GHz spectrum, targeting five common threat bands used by drones. With a detection range of approximately 9.66 km (six miles), the AUDS system employs micro-Doppler radar, high-precision infrared and daylight cameras, and advanced video tracking software to detect, locate, and confirm the presence of drones. Subsequently, it employs directional high-power RF interference to disrupt the communication link between the drone and its remote control, effectively forcing the drone to land. Remarkably, this entire process of detection, tracking, identification, and forced landing occurs within a rapid 15-second timeframe. The AUDS system has been successfully deployed by esteemed organizations like the British Army and the Metropolitan Police in the UK. Furthermore, it has gained international recognition with deployments in the United States and the Middle East, safeguarding critical infrastructures such as airports, power plants, and government buildings, as well as supporting military operations.

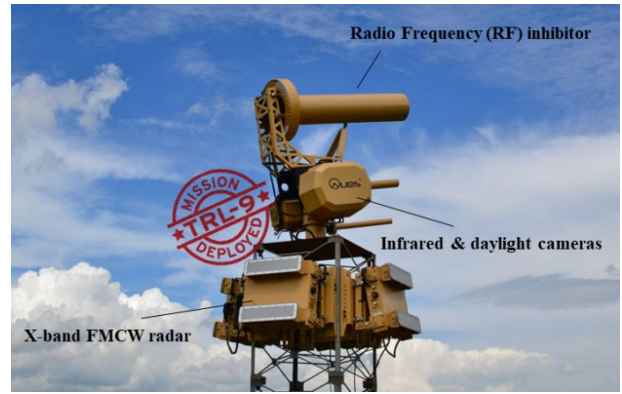

**(a)**

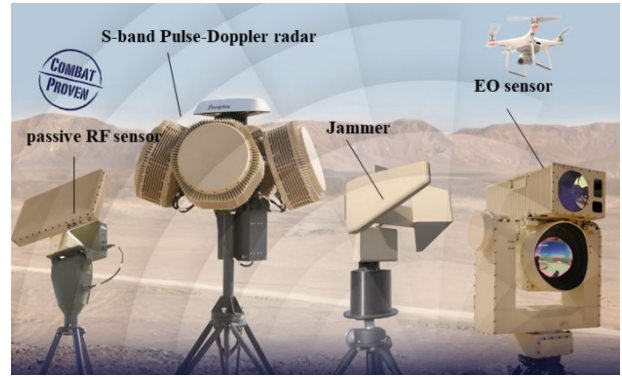

**(b)**

**Fig. 1.** Radar is the core detection sensors of C-UAS system. (a) the AUDS solution[1]. (b) Drone Dome system[2].

1. https://www.auds.com/auds-counter-drone-system-enhanced-for-vehicle-deployment-and-to-defeat-swarm-attacks/, accessed on May 5, 2023.
2. https://www.rafael.co.il/worlds/air-missile-defense/c-uas-counter-unmanned-aircraft-systems/, accessed on May 5, 2023.

Another notable C-UAS system is the Drone Dome, developed by Rafael, an Israeli company. Introduced in April 2016 at the Brazil Defense Exhibition, the Drone Dome system offers a comprehensive set of capabilities for countering drone threats. It incorporates the PRS-42 tactical airborne radar, the MEOS electro-optical sensor, and the C-Guard broadband signal jammer. The system's radar, in conjunction with Micro-Doppler classification and EO/IR sensors, enables the detection and identification of enemy drones. Upon detection, the system correlates and analyzes the collected data, issuing warnings to enemy drones. In scenarios where warnings are ignored and drones enter the designated kill zone, the operator can employ either hard kill or soft kill methods to neutralize the threat. Recent reports indicate that the Drone Dome system can be further enhanced with a high-energy laser, referred to as the Drone Dome-L, thereby providing an additional capability to destroy targets. With its ability to operate 24/7 under diverse weather conditions, the Drone Dome system has been recognized for its efficacy. In response to the Gatwick Airport drone incident in 2018, the British Army procured six sets of Drone Dome anti-drone systems for £16 million. Additionally, the system has been acquired by Japan to enhance security measures during the Tokyo Olympics in 2021. Notably, the US Army's Joint Counter-Small Unmanned Aircraft Systems Office has recently approved and recommended the inclusion of Drone Dome anti-drone systems in the C-UAS as a Service whitelist, thereby enabling authorized provision of anti-drone services and hardware to the US government.

A general C-UAS solution encompasses three key components: (1) the detection unit, (2) the decision-making unit, and (3) the confrontation unit. These components work in a coordinated manner. The detection unit, primarily utilizing radar technology, is responsible for the initial detection and recognition of potential threats. Concurrently, electro-optical/infrared systems validate and confirm the identified targets. Subsequently, the decision-making unit processes the detection information and selects appropriate countermeasures tailored to the specific type of target. Finally, the confrontation unit executes the chosen countermeasures to neutralize the unmanned aircraft threat.

Radar technology holds a pivotal role in C-UAS solutions, providing distinct advantages such as active detection capabilities, long-range coverage, all-weather functionality, detectability, and trackability. Radar performs two fundamental functions within the C-UAS system. Firstly, it scans the designated area, enabling early detection and providing crucial information to the decision-making unit. Additionally, radar facilitates precise target measurement, guiding the confrontation unit in executing effective countermeasures. As a result, radar serves as a cornerstone sensor in C-UAS solutions, contributing significantly to their operational effectiveness.

## II. DRONE DETECTION RADAR SYSTEMS

Traditional air defense radars and air traffic control radars are not well-suited for detecting small drones. These radar systems, such as the Russian S300 air defense system, primarily focus on high-speed large aircraft and ballistic missiles, rendering them inadequate for effectively detecting and tracking small drones. While these systems excel in detecting, tracking, and engaging fast-moving and large targets, they are often inefficient when it comes to dealing with small, slow-moving, and low-altitude Drones. Attempting to adapt traditional air defense radars to address the specific challenges posed by small Drones would be akin to using excessive force for a minor task.

Traditional radars used in air defense applications employ strategies to minimize interference from small and slow-moving targets. These measures include raising the beam elevation angle to prioritize high-altitude targets, raising the speed threshold to focus on fast-moving targets, and increasing the signal detection amplitude threshold to emphasize larger targets with strong scattering echoes. Consequently, these adjustments inadvertently disregard small Drones, which are constructed from composite materials and exhibit temperatures that closely resemble the surrounding atmosphere. Their relatively low speeds, approximately 150 kilometers per hour, further contribute to the challenges of detection, as they are comparable to the speed of clouds. Consequently, when air defense systems attempt to eliminate cloud interference, they inadvertently filter out low-speed targets like Drones, as these targets are not considered within the intended scope of the radar's design. Such filtering is essential to prevent false alarms caused by various non-threatening objects, such as birds. To address the shortcomings of traditional radar systems in detecting Drones effectively, the development of dedicated drone detection radars is imperative. These specialized radars should be designed to account for the unique characteristics of small, slow-moving targets, enabling reliable detection and tracking of drones in diverse operational environments.

Many countries worldwide are actively advancing the development, procurement, and deployment of anti-drone radar systems to tap into the expanding anti-drone market. While variations exist in terms of radar frequency bands, system designs, and other technical aspects, these systems generally adhere to fundamental radar principles. Table II provides an overview of some prominent drone detection radar systems available in the market. Despite variations in radar parameters, their overall detection performance remains largely consistent. The X-band radar band emerges as the most widely used across these systems. Additionally, micro-Doppler analysis combined with kinetic features serves as the primary basis for Automatic Target Recognition (ATR) of drones. Moreover, these radar systems demonstrate a detection range of up to 6 km for drones with a RCS of approximately 0.01-0.1 $m^2$, as observed in the case of the DJI Phantom drone.

TABLE II Some commercial drone detection radars[1]

| Model | Country | Band | Range (km)[1] | Possible identification features |
|---|---|---|---|---|
| Meteksan - Retinar FAR-AD | Turkey | Ku | 4.4 | Micro-Doppler |
| HENSOLDT- Spexer 2000 3DMKII | Germany | X | 6 | Micro-Doppler, Trace, etc. |
| AVEILLANT- Gamekeeper 16U | UK | L | 5 | Micro-Doppler, Trace, etc. |
| QinetiQ- Obsidian | UK | X | 2 | Micro-Doppler, etc. |
| Blighter-A800 | UK | Ku | 3 | Micro-Doppler, etc. |
| Weibel - XENTA-M1 | Denmark | X | 10 | Micro-Doppler, etc. |
| Retia - ReGUARD | Czech | X | 6 | RCS, etc. |
| ART-ART Midrange 3D | Spain | X | 3.6 | Micro-Doppler, etc. |
| IAI- ELM/2026BF | Israel | X | 5.2 | Trace, etc. |
| Elbit system- DAiR™ | Israel | X | 6 | Trace, micro-Doppler, etc. |
| Robin- ELVIRA | Netherlands | X | 2.7 | Micro-Doppler, etc. |
| Thales-GO20 MM | France | X | 4 | Micro-Doppler, etc. |
| Thales- SQUIRE | France | X | 6 | Micro-Doppler, etc. |
| SAAB-Giraffe 1X | Sweden | X | 4 | Trace, Micro Doppler, etc. |
| Leonardo DRS-RPS-42 | Italy/U.S. | S | 10 | Micro-Doppler, etc. |
| Teledyne FLIR- R20SS-3D | U.S. | X | 4 | NA |
| Raytheon-KuRFS | U.S. | Ku | NA | Image, Trace, etc. |

1. These data can be found in their official websites.
2. The update rate is the typical value. Some of them can be selectable.
3. The detection range is for drones with RCS of ~ 0.01 $m^2$, like DJI Phantom-4. The classification range is normally shorter than the detection range.
4. The specific identification signatures are not available, and those terms are reported in their official brochures.

In a realistic scenario, the effectiveness of a drone detection radar relies on its ability to distinguish radar signals emitted by drones within a complex background. Designed primarily for monitoring airspace below 1000 meters Above Ground Level (AGL), the drone detection radar is capable of detecting clutter from both ground-based and upper air objects (Fig. 2). Some drone detection radars attempt to mitigate ground clutter by adjusting the radar's elevation angle. However, this approach poses the risk of overlooking potential targets since most small drones typically operate within the super-low-altitude airspace below 100 meters. Consequently, it is important to recognize that a drone detection radar primarily serves as a surface surveillance radar rather than an air defense radar.

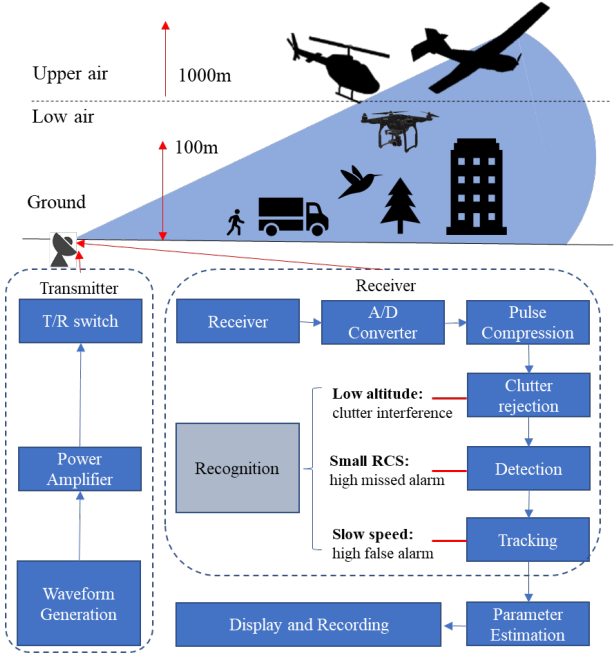

**Fig. 2.** The detection and recognition problem of radar detecting drones with the signal processing chain in the scenario of detecting drones.

The utilization of drone detection radar systems poses various challenges that affect the radar's signal processing steps, as depicted in Fig. 2. In this context, we will consider the example of a pulse-Doppler radar. Firstly, the "Low, Small, and Slow" (LSS) characteristics of drones significantly influence radar system operations. Specifically, the "low altitude" of drones necessitates the drone detection radar system to contend with clutter from ground backgrounds and moving objects, such as people, vehicles, and notably, birds. Consequently, an effective drone detection radar system must possess ATR capabilities. Secondly, the "small size" of drones requires the radar detector to exhibit sufficient agility in order to extract weak signals and attain an extensive detection range. This prerequisite demands a high level of recognition confidence. Thirdly, the "slow speed" of drones poses challenges for radar tracking, potentially resulting in misguidance of the confrontation unit within the C-UAS solution. This situation necessitates a high update rate for radar scanning and fast recognition speeds.

In summary, a reliable and efficient drone detection radar system requires a robust ATR function. By addressing these challenges and incorporating the appropriate capabilities, drone detection radar systems can enhance their performance and contribute significantly to the field of counter-drone operations.

## III. AUTOMATIC TARGET RECOGNITION (ATR)

ATR technology is widely recognized as being at the forefront of radar advancements. However, due to its sensitive nature, many technical details pertaining to ATR are not publicly available. Yet, many radar solutions own the ATR functions. For instance, Fig. 3 illustrates a representative ATR solution offered by the SAAB Giraffe radar system. The SAAB Giraffe radar is a highly sophisticated system renowned for its capability to detect and track aerial targets. Specifically, its Enhanced Low, Slow and Small (ELSS) feature is tailored to identify and monitor low-flying, slow-moving objects such as unmanned aerial vehicles (Drones), cruise missiles, and other diminutive airborne entities.

The ELSS technology, as reported by the manufacturer, incorporates advanced signal processing techniques, high-resolution radar signatures including radar cross-section (RCS), kinetic features, micro-Dopplers, and 3D mapping capabilities. These features empower the system to detect and track targets even in complex and cluttered environments. Notably, the SAAB Giraffe radar can differentiate drones amidst a flock of birds, as depicted in Fig. 3a, where the green traces represent birds and the yellow traces represent drones. The recognition outcomes are then displayed on the radar screen, exemplified in Fig. 3b. The illustration in Fig. 3 emphasizes that birds present a significant challenge as clutter for drone detection, and the recognition process involves a certain level of probability. If we refer to the recognition tier, the SAAB radar may achieve the Tier **Identification.**

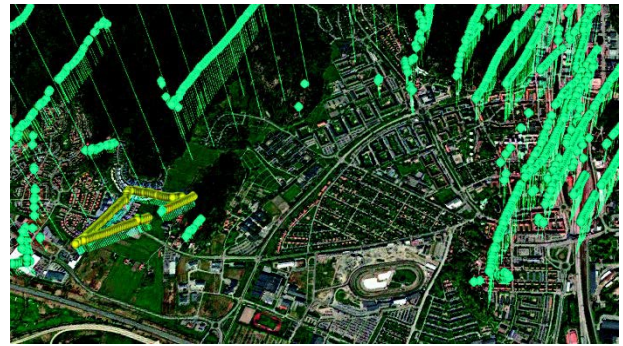

(a)

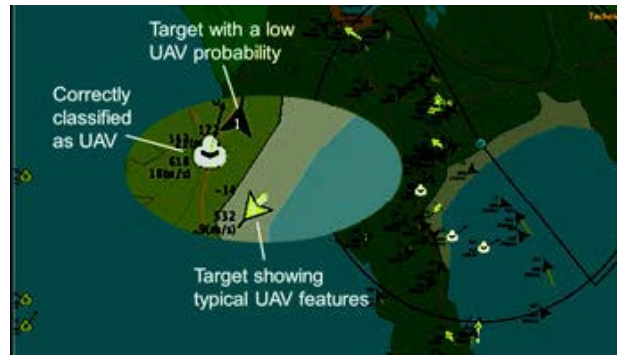

(b)

**Fig. 3** The detection & recognition results of drones provided by the SAAB giraffe radar[1], (a)multi-objects traces, (b) the recognition results.

1. https://www.saab.com/products/giraffe-1x, accessed on May 5, 2023.

## A. *Scattering regions*

In general, ATR function is typically regarded as an additional module that is integrated into an existing radar system. This implies that the ATR module operates within the constraints imposed by the radar parameters. Consequently, the performance of the ATR module may be limited by certain radar parameters that are not optimally suited for the specific ATR solution. Among these parameters, one of the most overlooked aspects is the radar band, or more precisely, the scattering regions associated with it, as shown in Fig. 4. The scattering of a target can be categorized into three distinct regions based on the ratio of its size to the radar wavelength: the Rayleigh region, the resonance region, and the optical region.

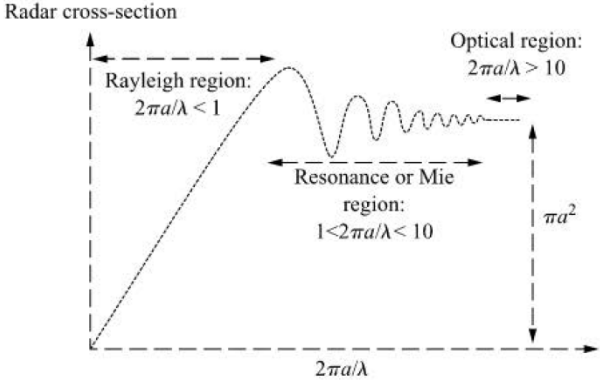

**Fig. 4** The variation of RCS of spheres within three scattering regions, where $a$ is the radius of the sphere, $\lambda$ is the wavelength [8][9]

(1) Rayleigh region:

Within this region, the target's size perpendicular to the wavefront is significantly smaller than the wavelength of the incident wave. Consequently, the incident wave does not undergo substantial phase changes when interacting with the target. The RCS of the target follows a linear relationship with frequency and can be approximated as a point source. Echo data obtained from this region provides only rudimentary information such as target size and volume.

(2) Resonance region:

In the resonance region, the target's size perpendicular to the wavefront is comparable to the incident wavelength. The scattering behavior of the target primarily arises from surface waves. The RCS of the target becomes a function of both the target size and the wavelength, leading to echo characteristics that exhibit alternating peaks and valleys due to interference between the scattering field components. The resonant region unveils the inherent frequency structure of the target, and a radar system employing multiple polarization states can provide a comprehensive description of the target's poles. Analyzing these poles enables determination of the natural frequency, and then the material composition and identification of the target type. Thereby, "pole" theory related to the natural frequency is basic in the resonance region.

(3) Optical region:

Within the optical region, the target's size perpendicular to the wavefront greatly exceeds the incident wavelength. The RCS of the target tends to remain relatively constant. Scattering primarily occurs through specular reflection and the field of the edge segment, which are determined by the strong scattering points present on the illuminated surface of the target. The total strength of the scattering field can be approximated as the sum of these strong scattering points. Radar echoes received from this region encompass detailed geometric and structural information of the target, rendering them valuable for automatic target recognition (ATR) purposes. "Scattering centers" theory related to the geometry is the basic in the optical region.

Different scattering regions necessitate the consideration of distinct ATR methods. Notably, Peter Trait has contributed significantly to the field of radar ATR and has authored a book addressing this subject, which outlines fundamental principles governing radar ATR solutions [10]. Similarly, David Blacknell and Hugh Griffiths have published a comprehensive book on radar ATR [11], delving into various aspects of target categorization, encompassing ground targets, air targets, and maritime targets. Over the course of several decades, multiple ATR schools have recognized the significance of radar features in achieving successful ATR applications in specific cases.

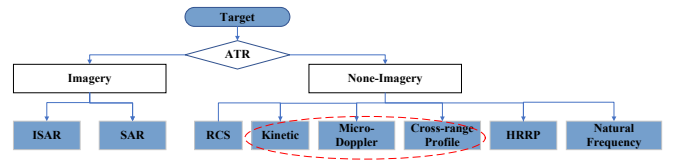

**Fig. 5.** The flowchart of designing a radar driving by different ATR methods (the red dotted circle marks the one for drone detection radar)

## B. *Recognition signatures*

In recent years, there have been significant advancements in radar ATR technologies, mainly due to the emergence of Deep Learning (DL) methods. However, it is important to note that DL methods primarily serve as a technique rather than signatures themselves. The ATR processing for radar images involves both images and non-images (Fig. 5). Radar imagery encompasses Synthetic Aperture Radar (SAR) images and Inverse Synthetic Aperture Radar (ISAR) images. SAR is commonly used in air-based radar systems, whereas ISAR is prevalent in sea-based radar systems. In this context, our focus is in ground-based radar systems.

Traditional ATR solutions typically employ a template-based matching approach that involves two key steps: feature extraction and pattern recognition. Consequently, the radar signatures play a crucial role in the initial stages. Based on the features extracted from these radar signatures, they can be categorized into two distinct groups: the long-term characteristics, and the transient ones.

(1) Radar Cross Section (RCS)

One category of features used in ATR is statistical features. Radar can detect a target and measure various values related to it, which can be divided into two parts: Radar Cross Section (RCS) related values and kinetic values. RCS, which represents the scattering power or electric size of the target, serves as a common signature for classification. The RCS of a target, $\sigma$ is given as:

$$\sigma = \lim_{r \to \infty} 4\pi r^2 \frac{E_s^{\ 2}}{E_i^{\ 2}} \qquad (1)$$

where, $r =$ the range, $E_s =$ the scattered field, and $E_i =$ the incident field. Thereby, RCS is the area intercepting that amount of power which, if radiated isotropically, produced the same received power in the radar. It is a statical mean value. For example, a jet usually has an RCS level of 100 m², while a small quad-rotor drone's RCS lies around 0.1 m². The measured RCS can fluctuate over time, and it can be represented as a waveform or time series. Additionally, the statistical features derived from RCS values can provide valuable information about the target, such as altitude, which can aid in target recognition.

### (2) Kinetic features

Another category of features is kinetic features, which include speed and trajectory. Consider the target's speed is $V_b$, which is determined though Doppler measurements in the radar system. The Doppler shift, can be given as:

$$\overline{f_{bd}} = -\frac{2V_b}{\lambda} \tag{2}$$

where, $\overline{f_{bd}}$ = the Doppler shift, and $\lambda$ =the radar wavelength. Subsequently, the trace function of the target, denoted as $L(t)$ can be defined as:

$$L(t) = V_b t \tag{3}$$

where, $t$= the measured time. The essential kinetic features encompass differential functions of the measured trace, taking into account factors such as velocity, time, and range. Mathematically, this can be represented as,

$$D(r,t) = f\left(\frac{dL}{dt}, \frac{dL}{dr}, \frac{dV_b}{dt} \dots\right) \tag{4}$$

where, $D(r,t)$ = the general kinetic features, and $L$ =the detection range of the target, $r$= the range resolution of the radar. These features, also referred to as trace classification, have long been employed in radar systems, especially for classifying aerial targets. When a radar system detects an aerial target, it generates a trace, which is a chronological record of the target's movement. Analyzing these traces using trace classification algorithms allows for the determination of various target characteristics, including size, speed, and flight pattern [12]. This information plays a crucial role in classifying the target as a drone, airplane, helicopter, or other types of aerial vehicles. However, there are certain considerations when utilizing kinetic features. Firstly, successful target detection is necessary for extracting kinetic features. Since drones typically emit weak radar signals, the detection probability significantly affects the performance of trace classification. Additionally, the presence of birds with trace patterns similar to those of drones poses additional challenges to ATR performance. As a result, trace classification requires longer processing time and has a shorter recognition range, as kinetic features do not exhibit robust signature characteristics.

### (3) The range-profiles

Geometry information can be explained using the scattering center theory in the optic region. The High Range Resolution Profile (HRRP) technology is widely used to separate the scattering centers of a target [13]. Radar systems transmit ultrawide-band signals and capture target profiles in the range direction. The HRRP provides a representation of the target's shape, which is then processed using template matching techniques with the dataset to determine the target class. The U.S. Air Force Research Laboratory (AFRL) conducted a notable project called the Systems-Oriented High Range Resolution (HRR) Automatic Recognition Program

(SHARP) [14]. In essence, the HRRP maps the target's shape information projected in the range domain, which is defined by the range resolution equation, given by

$$R_e = \frac{c}{2B} \tag{5}$$

where, $R_e$= the range resolution, $c$= the transmitted velocity of light, and $B$= the transmitted bandwidth.

### (4) The cross-range profiles

Since the Doppler resolution is directly proportional to the cross-range resolution, the cross-range profile of an object is also a separation of scattering centers of the target. The application is the ISAR technology [8], which is described by

$$\Delta f_d = \frac{2\Delta R_c \omega}{\lambda} \tag{6}$$

Where, $\Delta f_d$= the Doppler resolution, $\Delta R_c$=the distance of the two scatter centers, $\omega$ =the angular rotation rate, and $\lambda$ =the radar wavelength. The cross-range profile technology can also be used for classifying small targets. For example, it can be used for classifying radar echoes from vehicles and helicopters [15], or the small birds and large birds [16]. Cross-range resolution improves with shorter wavelengths and over larger rotation angles. It is interesting that the cross-range resolution is independent of the range of the target, unlike SAR, in which the synthetic aperture has to be increased at long ranges to maintain range resolution.

### (5) The micro-Doppler signature

The micro-Doppler approach can extract specific geometry from a target. Micro-Doppler refers to the additional Doppler component generated by micro-motions exhibited by the target, such as the rotational movement of helicopter blades or the flapping of bird wings [17]. Micro-Doppler analysis enables the extraction of embedded kinematic and structural information, which can be used for target registration. While micro-Doppler signatures are typically treated as kinetic features, in this context, we consider some specific micro-Doppler signatures as structure/geometry signatures. Dr. Chen highlights a key challenge associated with the micro-Doppler method, emphasizing the need for effective interpretation of the extracted features and their correlation with the structural aspects of the target under scrutiny [17]. In the context of drones, the micro-Doppler phenomenon primarily stems from the coherent rotational motion or the presence of rotating structures, as exemplified in Fig. 6b. Consequently, this phenomenon enables the extraction of distinctive radar signatures pertaining to drones, such as the number and rotational rate of blades, which greatly facilitate drone recognition. Consequently, as depicted in Table II, a multitude of drone detection radar systems employ micro-Doppler analysis for the identification of drone-related radar signals. The observed micro-Doppler effect is also a consequence of the classic Doppler effect. Assuming a drone's velocity as $V_b$, the micro-Doppler shift can be approximated by considering the radial component of the blade velocity projected onto the radar's line-of-sight. The relationship can be expressed as:

$$\overline{f_{md}}(t) = \frac{L}{\lambda} \omega \cos\alpha \cos\beta \cos(\omega t) + \overline{f_{bd}} \tag{7}$$

where, $R$ = the radar range, $L$= the blade length, $\omega$= the rotating rate of blades, $\alpha$= the azimuth angle, and $\beta$= the elevation angle. Based on the aforementioned equation, the micro-Doppler shift is modulated by two sinusoidal functions. Consequently, a comprehensive description of micro-Doppler signatures encompasses the modulation function, which is inherently dependent on the observation time.

(6) The natural frequency

The natural frequency of a target is a function of its shape and material contents. Since the natural frequency is independent of range and altitude, it exhibits robustness that can enhance target recognition [18][19][20][21]. It is important to note that utilizing this method requires the basic pre-condition of understanding the scattering polar theory in the resonance region with the "poles" theory. However, extracting the natural frequency and linking it to specific objects is challenging, and practical radars utilizing this method are rare in the market. Most of the research on this topic remains in laboratories due to the lack of clarity surrounding the scattering mechanism in the resonance region.

## C. Drone ATR methodologies

As for drone detection radar applications, it is worth noting that the detection and identification of small drones using high range resolution radar profiles (HRRPs) pose challenges due to the need for sub-centimeter resolution to capture the longitudinal structure of targets measuring less than 100 cm in length [22][23]. Consequently, many radar systems employed for bird and drone detection utilize low range profiles, making micro-Doppler and kinetic features the primary choices for identification purposes.

However, both kinetic features and micro-Doppler have limitations when it comes to drone automatic target recognition (ATR). The main issue with kinetic features is their vulnerability to variations in time. As shown in equation (4) above, because range and velocity are influenced by the sampling time, the resulting kinetic features lack robustness and fail to provide distinctive signatures. On the other hand, micro-Doppler has become increasingly popular in the drone recognition field. Equation (7) emphasizes the importance of observation time for effectively utilizing micro-Doppler.

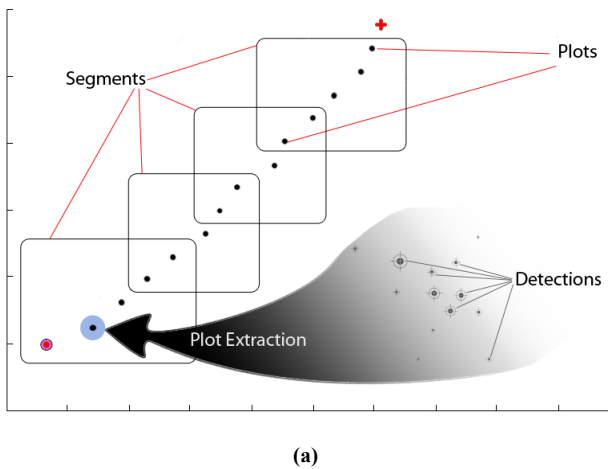

(a)

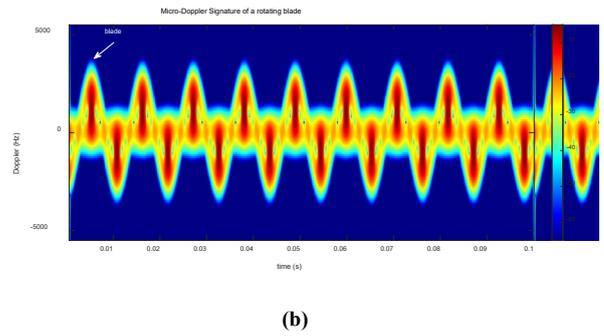

(b)

Fig. 6. The two recognition method examples, (a) The example of segments with 5 plots and 2 overlapping plots trace classification [24], (b) The simulated micro-Doppler spectrogram of one blade[25].

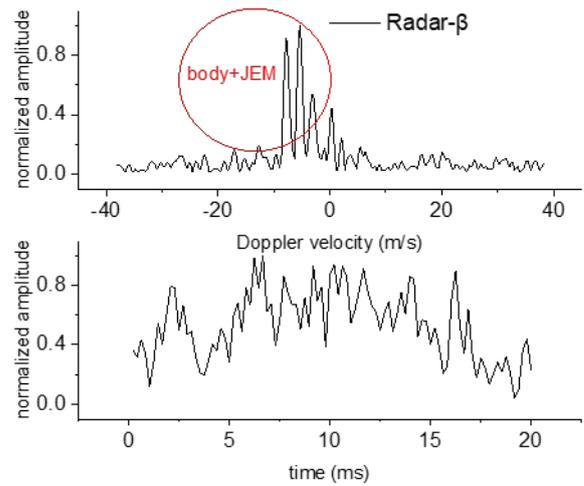

(a)

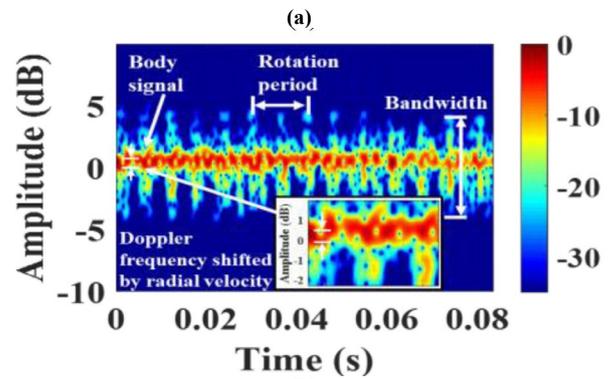

(b)

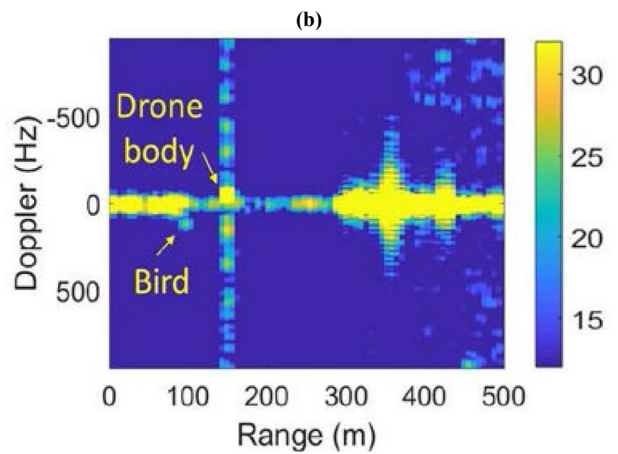

(c)

**Fig. 7.** The prestation forms of micro-Doppler of quad-rotor drones, **(a)** the JEM-like spectrums detected by a X-band pulse-Doppler radar [26], **(b)** the "blade flash" patterned spectrogram detected by a X-band radar [27], **(c)**The micro-Doppler range-Doppler detected by a X-band FMCW radar [28].

Micro-Doppler classification, despite its advantages, also faces certain challenges. Firstly, micro-Doppler is more of a phenomenon than a signature. Fig. 7 illustrates three forms of the micro-Doppler phenomenon: Jet Engine Modulation (JEM) or Helicopter Rotor Modulation (HERM) lines modulation in the spectrum, the 'blade flash' patterned spectrogram obtained through the Short-time Fourier Transform (STFT), and the range-micro-Doppler profile. JEM spectrum refers to spectral peaks with certain adjacent intervals and similar amplitudes in the spectrum [15], while the 'blade flash' pattern describes the sinusoidal trace on the spectrogram. The instant micro-Doppler signature represents the stable mapping of scattering centers of rotating blades in the cross-range profile, capturing the scattering characteristics or structures of the rotating component rather than the rotational pattern. Quantifying the micro-Doppler signature poses a challenge that needs to be addressed. Secondly, enhancing micro-Doppler signals comes at a cost. Not all radar dwell times are suitable for detecting micro-Doppler signals produced by the rotating blades in drone radar signals. The radar dwell time should be neither too long nor too short for effective micro-Doppler detection [25]. Additionally, a sufficiently high sampling frequency is required to obtain an adequate amount of micro-Doppler information for separating the micro-Doppler signals. Insufficient micro-Doppler data can lead to incorrect estimates of blade numbers [29]. Moreover, achieving a high signal-to-noise ratio (SNR) of micro-Doppler images using machine learning algorithms is essential for accurate drone recognition [30]. In summary, the micro-Doppler method has a short detection range and recognition range but a long detection response time (DRT). Despite these challenges, micro-Doppler recognition remains a crucial component of modern radar systems, enabling the accurate identification and classification of small aerial targets like drones. With the increasing use of drones, the importance of micro-Doppler recognition in drone detection and monitoring systems is expected to grow.

## IV. GUIDING PRINCIPLES BY ATR AND CASE STUDIES

### A. *The design principles*

The design of traditional radar systems is guided by the radar equation, which assumes that a target can be represented as a point object with a mean RCS. This equation allows for the calculation of the signal-to-noise ratio (SNR), a measure of a radar system's ability to detect a specific target at a given range, by comparing the target's scattering power with the background noise. The radar equation is mathematically expressed as follows [31]:

$$R = \sqrt[4]{\frac{P_t G_r G_t \lambda^2 \sigma}{(4\pi)^3 K T_s B_n L(SNR)}} \qquad (8)$$

where $T_s$ = the system noise temperature, $B_n$ = the noise bandwidth of the receiver, $L$ = the total system losses, $K$ = the Boltzmann's constant. $P_t$ = the transmitted power, $G_r$ = the received gain, $G_t$ = the transmitted gain, $R$ = the measured range, $\sigma$ = the RCS of the target, $\lambda$= the radar wavelength. The radar equation illustrates that the fundamental rule of radar design is based on radar detection principles. For single pulse detection, if the detection probability exceeds 50%, the SNR of the target should be at least 13.1 dB. To achieve a 95% probability of detection, the SNR should be 16.8 dB [31]. In simpler terms, smaller targets with lower RCS values will have lower SNRs, resulting in a reduced detection range. Consequently, radar systems often encounter the challenge of "Missed Target" when detecting drones and other objects with small RCS values.

The design of drone detection radar systems should be guided by the ATR method. Fig. 5 provides a comparison between two types of drone detection radars that utilize different ATR methods. In this discussion, we will use the pulse-Doppler radar as an example.

(1) The ATR method based on kinetic features primarily addresses the detection problem.

To increase the recognition probability, it is beneficial to improve the detection probability. Firstly, the resonance region is crucial for kinetic features as it amplifies the RCS of the target and enhances the detection probability. Given that small drones typically fall within the submeter size range, S-band or L-band radars are more suitable for detecting small drones and achieving higher detection probabilities. Moreover, a higher detection probability contributes to better radar tracking, which, in turn, improves ATR using kinetic features. Secondly, increasing the efficiency of ATR can be achieved by accelerating the tracking rate to enhance the kinetic features. Kinetic features are characterized by time and range variations. Therefore, a faster inspection rate translates to a higher update rate of kinetic features but requires shorter radar dwell time.

(2) The ATR method based on signal signatures addresses the classification problem.

To increase the recognition probability, it is preferable to enhance the signal signatures. Taking the micro-Doppler signature as an example, although it falls under the category of kinetic features, it also captures specific target structures. The micro-Doppler "flash" or JEM spectra reflect the rotating blades of drones, which can be utilized for ATR. Firstly, the optical region is more favorable than the resonance region. The micro-structures are attached structures on the target body and are smaller in size compared to the body itself. Consequently, the resonance effect in the resonance region may amplify the scattering power of the body but suppress the micro-Doppler scattering power. Additionally, a shorter wavelength results in a better Doppler shift and a larger difference between the body Doppler shift and the micro-Doppler shifts. As illustrated in Fig. 8a, L-band radar offers higher SNR values (over 20 dB) compared to X-band radar, but its JEM signature is weaker. Conversely, X-band radar can detect almost 100% of JEM spectra, making it more effective in drone detection. Secondly, since micro-Doppler is a result of the Doppler effect, a higher Doppler resolution enhances the micro-Doppler signatures. Therefore, longer radar dwell times and higher frequency resolutions are favorable for achieving better Doppler resolution. In most cases, this entails a slower inspection rate or tracking rate. As depicted in Fig. 8b, there exists an optimal radar dwell time for micro-Doppler of drones. In our analysis, we examined radar data detected by three radar systems with different radar dwell times but similar frequency and velocity resolutions, including Radar-α and Radar-γ, with radar dwell times of 2.7 ms and 89 ms, respectively. Radar-α barely detected any micro-Doppler, while Radar-γ captured weak micro-Doppler signals, with a magnitude only 10% of the body Doppler's. Proper radar dwell time is crucial for micro-Doppler detection. This research provides insight into designing a cognitive micro-Doppler radar by adjusting radar dwell time to detect and track micro-Doppler signals of drones.

In summary, when designing a drone detection radar system guided by the ATR method, it is important to consider not only the traditional detection unit but also the recognition unit. Several key aspects should be taken into account. Firstly, careful selection of the radar wavelength is essential due to the influence of scattering regions. The choice of wavelength should be made with consideration of the specific scattering characteristics of the target being detected. Secondly, the transmitted radar parameters, including radar dwell time (Coherent Processing Interval or CPI), sampling rate (Pulse Repetition Frequency or PRF), and others, play a crucial role in enhancing both the kinetic features and signal signatures. Optimal values for these parameters can improve the detection and classification capabilities of the radar system. Thirdly, the tracking strategy must be carefully considered within the ATR method. It is important to note that a higher tracking rate corresponds to a shorter radar dwell time. However, a shorter dwell time results in a lower SNR, which can potentially lower the detection probability. This contradiction between the detection and tracking units must be carefully managed, as it ultimately impacts the performance of the ATR unit. Therefore, it is imperative to adopt a comprehensive ATR approach when designing a drone detection radar system, taking into account all these factors and striking a balance between the conflicting requirements of detection and tracking. By doing so, the effectiveness and efficiency of the radar system can be maximized for accurate and reliable drone detection.

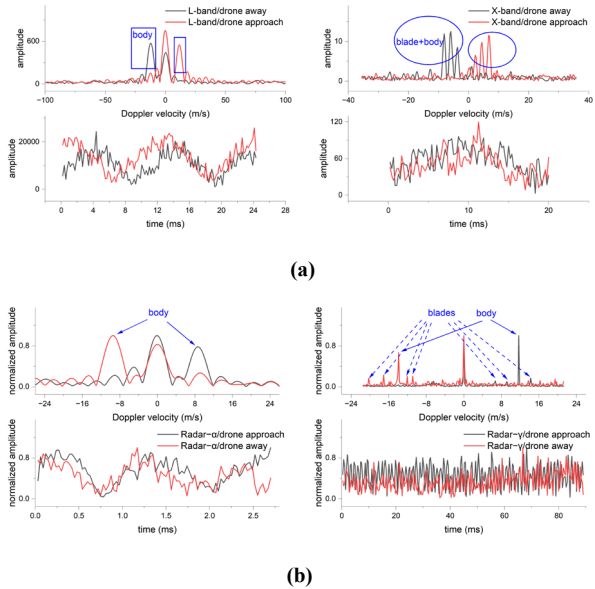

**Fig. 8.** The examples of micro-Doppler of quad-rotor drone, **(a)** the comparison detection using L-band and X-band radar, **(b)** the comparison detection using different X-band radar with different CPIs, where Radar−α, and Radar−γ with radar dwell times of 2.7 ms, and 89 ms, respectively [25].

### B. The drone detection radar systems with ATR function

In this section, we present our drone detection radar system, referred to as the WHU system, along with the results obtained using the ATR function. Fig. 9 illustrates the diverse clutter backgrounds encountered by our X-band drone detection radar, and the performance metrics are summarized in Table III. The radar operates at a frequency of approximately 9 GHz, with a CPI of about 20 ms and a PRF of 5 kHz. The transmitted band encompasses approximately 12.5 MHz, providing a range resolution of approximately 12 m. Equipped with an active electronically scanned phased array (AESA) antenna, the radar system is mounted on a rotating table to achieve a 360-degree coverage in azimuthal scanning. A comparative analysis of the detection performance reveals that our radar system exhibits an extended range for drone detection, faster recognition speed, and higher recognition accuracy, which can be attributed to the implementation of the ATR function.

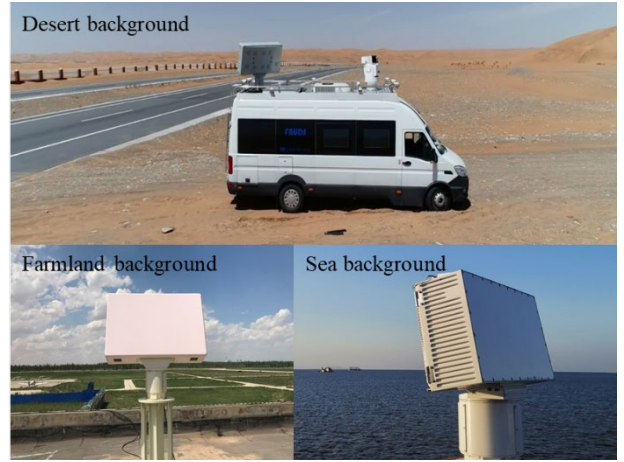

**Fig. 9.** The example of our drone detection radar (WHU one) in the different clutter backgrounds.

TABLE III The comparison of drone detection radars

| Contents | General ones[1] | WHU one |
|---|---|---|
| Systems | FMCW, Pulse-Doppler | Pulse-Doppler |
| Bands | S, L, X, Ku, etc | X, Ku, etc |
| Antenna | Phased array antennas | Phased array antennas |
| Scattering region | Resonance region, Optic region | Optic region |
| Detection method | SNR based | SNR & SCR based |
| Recognition method | Micro-Doppler, trace, imaging, etc. | Scattering & Doppler, etc. |
| Detection range[2] | < 6 km | > 12 km |
| Recognition range | < 3 km | > 12 km |
| Detection response time (DRT)[3] | ~1 s | < 10 ms |
| Data Inspection Frequency (DIF)[4] | Adjustable | Adjustable |
| Automatic Target Recognition (ATR) | Birds, Drones, Pedestrians, Vehicles, clutters. | Birds (Large birds, Small Birds), Drones(Fixed-wing drones, Multi-rotor drones, VTOL drones), Vehicles, Ships, Pedestrians, Helicopters, Jets, etc. |

1. The general ones refer to the most typical drone detection radars in the market, such as thoses in Table I.

2. Here, the detection range and recognition range means the one for comon quad-rotor drone, DJI phantom series.

3. The DRT represents the lags between an echo return, detection, and eventual display.

4. The drone detection radar systems must be capable of continuous operation to allow for the continuous inspection of the surveillance areas, and the DIF means the time interval between successive updates for the position of a tracked object.

Several factors contribute to the enhanced performance of our radar system. Firstly, being a pulse-Doppler radar, it is capable of detecting moving targets while effectively suppressing static clutter present in the background. The inherent raw velocity resolution of 0.75 m/s enables the detection of drones even at very low speeds. Secondly, operating in the X-band frequency range ensures that the radar scattering data from small drones originates from the optical region, facilitating the extraction of shape information from the radar signatures. Thirdly, the adoption of a narrow band and a range resolution of 12 m ensures that small drones cannot simultaneously occupy more than three range bins, thereby mitigating range migration issues. Fourthly, using both the traditional SNR detector and our newly developed Dynamic Signal-to-Clutter Ratio (DSCR) detector [32][33], the detection range for small drones with RCS levels ranging from 0.01 to 0.1 m^2 extends up to 12 km. Lastly, leveraging phased-array technology, our radar system has the capacity to simultaneously track up to 1000 targets. It can accurately classify and identify various targets, including birds, drones, vehicles, ships, humans, helicopters, among others, and present the detection results in a visual manner through graphical icons (Fig. 9). The tracking numbers associated with each icon provide a clear tracking reference.

The utilization of the ATR function can significantly enhance the performance of radar systems in both the detection and tracking units. The effectiveness of this function is exemplified in Fig. 10, which showcases the radar display cases referenced from our previous publications. The experimental tests were conducted in the coastal region of Qidong city, situated along the Yellow Sea coast in China. This area is characterized by a cluttered sea environment. The radar system was installed on the rooftop of a 12-meter tall building, enabling horizontal scanning of the sea surface. Initially, our project aimed to develop a prototype drone detection radar. Throughout the project, an infrared sensor and an optical camera were incorporated to support and validate the recognition results obtained from the ATR function. The project spanned several months in 2020, and this paper presents some of the data extracted from these experiments. The test area exhibited a sea scale ranging from Degree 3 to Degree 5. With Degree 3, the height of the waves ranged from 0.5 to 1.25 meters, while with Degree 5, it reached 2.50 to 4.00 meters.

During the tests, we collected radar signals from different types of drones, including a quad-rotor drone, a fixed-wing drone, and a hybrid Vertical Take-off and Landing (VTOL) fixed-wing drone. These drones were considered cooperative targets, and their key parameters can be found in our earlier publications [34][35]. The Albatross 1 is a homemade fixed-wing drone with dimensions of 1.08 m × 0.80 m and a mass of 0.3 kg. The DJI Phantom 4 is a well-known quad-rotor drone equipped with four lifting blades, measuring 0.40 m × 0.40 m and weighing 1.38 kg. Finally, the TX25A is a large hybrid VTOL fixed-wing drone featuring one pusher blade and four lifting blades, with dimensions of 3.6 m × 1.97 m and a mass of 26 kg. Fig. 10 displays images of these drones. Referring to Table I, both the Albatross 1 and DJI Phantom 4 belong to Group 1, while the TX25A belongs to Group 2. These drones represent the primary targets for the current C-UAS solution. Additionally, local fishing ships and birds were also included as test targets, and their radar data was collected for analysis.

(1) The Beyond-Range increased by the ATR function

The ATR function offers a remarkable capability to extend the detection range of radar systems, surpassing their theoretical limits. The range of a target can be theoretically determined using the radar equation, which considers the target's Radar Cross Section (RCS) value and Signal-to-Noise Ratio (SNR). This relationship can be expressed as follows: The function can be given as:

$$R \propto \sqrt[4]{\frac{\sigma}{SNR}} \qquad (9)$$

Thus, for instance, if a radar system can detect an aircraft with an RCS of 100 $m^2$ at a range of 60 km, it may only detect a small drone with an RCS of 0.01 $m^2$ at a range of 6 km. To achieve a 3 dB increase in the drone's detection range, the SNR must be reduced by 12 dB while keeping other factors constant. However, if the radar's SNR threshold is decreased from 16.8 dB [31] to 4.8 dB, the detection process will encounter a significant increase in false alarms, resulting in clutter that hampers the radar display [33]. Fig. 10a illustrates the excessive number of false alarms on the radar screen, with only a few large ships being detected.

In contrast, by employing the ATR function to process these false alarms, numerous clutter objects can be rejected, enabling the detection of both small and large objects within the area. Methodologically, the target identification provided by the ATR function is fed back to the detection unit, integrating detection with recognition (IDR). Fig. 10b demonstrates the detection and recognition of various targets, including ships, birds, drones, and others [34] [35]. Moreover, the red sector represents the theoretically expected range of 14 km for drones. However, the ATR function can achieve a Beyond-Range ability for the radar system, surpassing the theoretical limits. For instance, it can detect and track birds at ranges exceeding 16 km. In summary, the ATR function enables the radar system to extend the detection range of a target and even exhibit a Beyond-Range capability, as long as the ATR function is operational.

(2) The sub-class of targets recognized by the ATR function

When the geometry information of the target is extracted from the radar echoes, the sub-class of the targets can be recognized by the ATR function. Firstly, we propose using flight morphology for classifying radar signals of large birds and small birds, in that large birds habitually carry their feet stretched out behind them during flight, leaving their feet apart from the body and being recognizable to both human observation and radar detection, while small birds tend to carry their feet drawn up in front, clinging to the body, thus hiding their feet from detection [16]. The visibility of a bird's feet will register radar signatures on bird echoes, and these signatures contribute to the classification of radar echoes from different sized birds in relatively short data sampling time. And then, we categorize small drones into three types based on their blade types: fixed-wing drones with only puller blades, multi-rotor drones with only lifting blades, and hybrid vertical take-off and landing (VTOL) fixed-wing drones with both lifting and puller blades (Fig. 10b) [35], and the micro-Doppler signals of the puller blades were weaker and more stable than that of the lifting blades; thus, the detailed micro-Doppler signatures modulated by different blades can be used for improving drone detection and identification of drone types by drone detection radar.

(3) The Situation-Awareness enhanced by the ATR function

The ATR function plays a crucial role in enhancing the situational awareness of radar systems. The recognition of target attributes by the ATR function can be effectively correlated with tracking information, enabling the acquisition of comprehensive situational awareness in the radar monitoring area. The classify-while-scan (CWS) technology [36] employed in this context processes the raw data within a single radar resolution cell, leading to the identification of objects. The obtained target ID can serve various purposes, including recording, display on the radar screen, and assisting the tracking unit. By connecting the same ID across consecutive tracking data, a track-after-identify (TAI) process is achieved, facilitating seamless tracking of targets. Consequently, the CWS function processes radar data in each radar cell, providing outputs of targets along with the corresponding range cells in the radar beam. The continuous scanning of the radar beam captures the movements and trajectories of active targets, thus presenting a comprehensive situational awareness of the entire scenario on the radar display [36].

Fig. 10c illustrates a typical application of situational awareness, where birds are seen chasing ships to seize fish that have been disturbed by the ship's propellers. The white dotted lines depict the tracking traces of objects, while the numbers around the icons represent the tracking numbers of the objects. Specifically, the tracking birds with No. 6264 are observed flying around the ship with No. 6283, resulting in radar echoes from the birds interfering with the detection of the ship. The recognition process sometimes categorizes the data as either birds or the ship when they appear in the same radar cell. The birds habitually chase the ship in the sea to feed on the fish, which are stirred up by the rotating propellers. Thus, this situational awareness reveals an interesting behavior of sea birds. The detection response time (DRT) is approximately a0 ms, indicating that the lag between echo return, detection, and eventual display is only 10 ms. Consequently, situational awareness can be continuously updated in real-time, enhancing the radar system's capabilities to function as a WYSIWYG (What You See Is What You Get) system or a real-time sense-and-alert system.

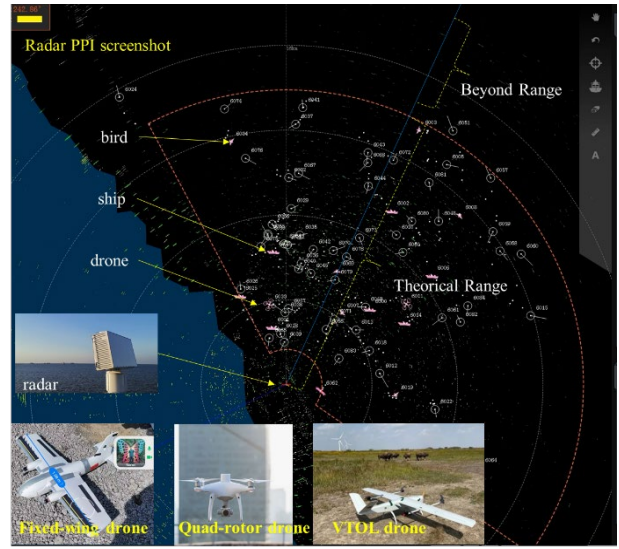

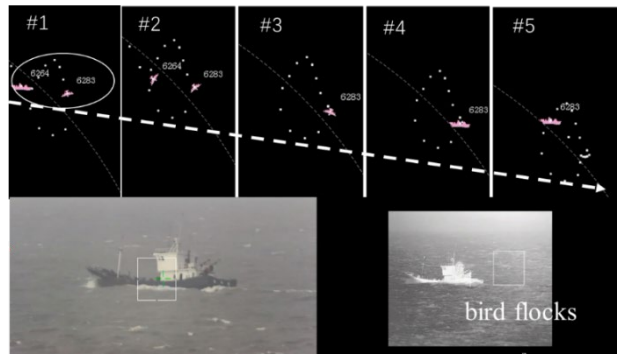

**(b)**

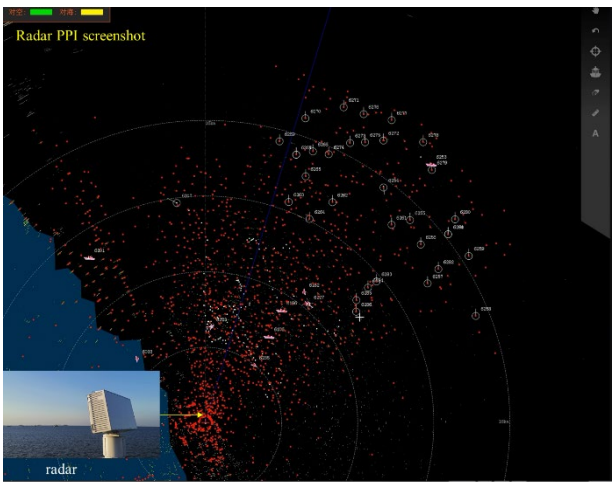

**(c)**

**Fig. 10.** The example of a drone detection radar, (a) a static detection screen short containing plenty of clutter, (b) a static detection display after using ATR function [35], (c) a dynamic Situation Awareness "video" enhanced by the ATR function [36].

*C.  Lessons learned from real-world implementations*

In recent times, numerous C-UAS solutions and drone detection radar systems have emerged, each claiming to possess exceptional performance in detecting radar signals emitted by drones. Some of these systems have been procured by clients and deployed in critical facilities. However, despite these advancements, certain governments remain skeptical about the functionality and value of drone detection radar systems. Consequently, they have initiated projects aimed at validating the effectiveness of such systems. In this context, we employ semi-simulated implementations to shed light on the significance of an ATR function in both C-UAS solutions and drone detection radar systems.

**Civil lessons**

**(a)**

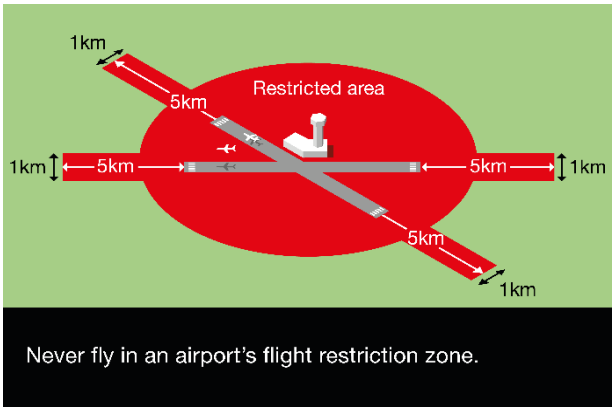

**(a)**

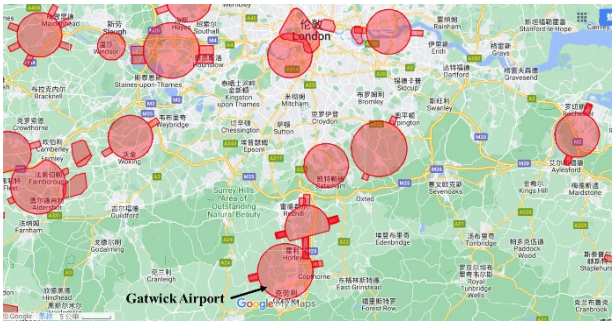

**(b)**

**Fig. 11.** Drones are banned within 5km of all UK airports. (a) no drone rules[1], (b) Gatwick Airport on the map[2].

1. https://register-drones.caa.co.uk/drone-code/where-you-can-fly
2. https://www.google.com/maps/d/viewer?mid=1BktWMPYNuh6N5_IPngyq8jW80nAWXI8d&ll=51.3123544615843%2C-0.20958776619308672&z=10

On the evening of December 19, 2018, drones were reported flying near the runway at Gatwick Airport in London, UK. As a precaution, airport authorities suspended all flight operations, resulting in significant disruptions for over 140,000 passengers, and more than 1,000 flights were canceled, causing 36 hours of disruption. This incident highlighted the need for security measures and strategies to protect against potential drone threats at airports worldwide. In response, the UK government implemented several measures, such as no-drone airspace within a five-kilometer radius (Fig. 11) and the deployment of C-UAS solutions, including the Drone Dome system. However, a subsequent drone disruption occurred at Gatwick Airport in 2019, indicating that the Drone Dome system was not a foolproof solution and that other factors contributed to the incident. The drone operator may have found ways to evade the system's detection and jamming capabilities, or multiple operators were involved, making it difficult to locate and neutralize all threats.

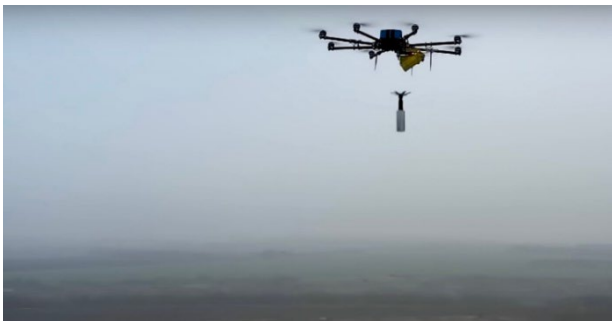

**(a)**

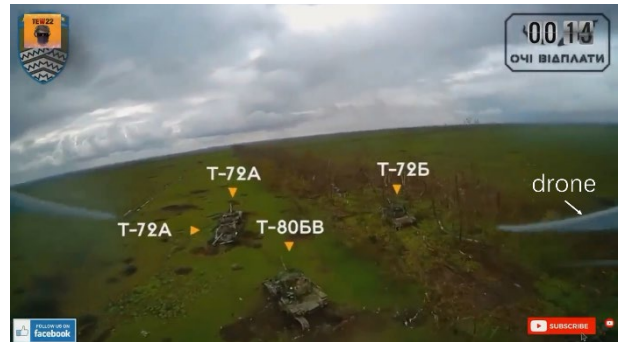

**(b)**

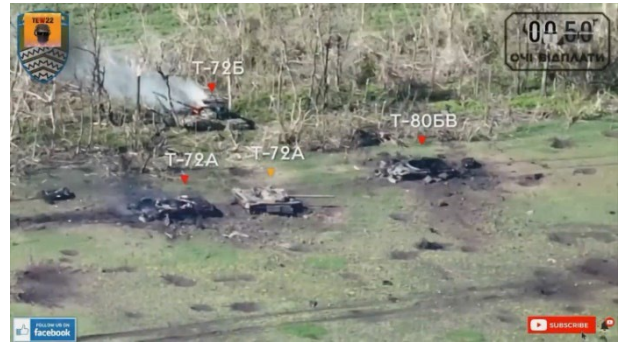

**(c)**

**Fig. 12.** Ukraine's forces drone eliminated 3 Russian tanks T-72A T-80B T-72B & Infantry Vehicle in South. (a) R18 UAV deployed in the ongoing war with Russia[1], (a) tanks tracked by drones in drone perspective[2], (b) destroyed tanks in scouting perspective[2].

1. https://dronedj.com/2022/09/16/aerorozvidka-ukraine-drone-r18/
2. https://www.youtube.com/watch?v=4PqOSWzI_x8

**Military lessons**

After Russia's invasion of Ukraine, the use of drones in combat has become increasingly prevalent and significant. Recent drone attacks by Ukrainian forces in the ongoing conflict have provided valuable insights and lessons. Drones equipped with advanced sensors, cameras, and precision targeting capabilities have proven to be highly effective in engaging and neutralizing heavily armored vehicles and infantry, which has leveled the playing field against Russian forces. Ukraine's use of drones exemplifies the principles of asymmetric warfare, by exploiting innovative tactics to challenge a stronger opponent while minimizing direct exposure to combat.

Civilian drones have transformed into multi-purpose assets, combing various combat capabilities. Micro-drones available in the market for less than $10,000 offer a wide range of possibilities, including obtaining information, preparing ambushes, designating targets for artillery, and tracking troop movements and fighter take-offs. Drones have provided Ukraine with a cost-effective means to challenge Russia's military superiority, disrupting their forces and undermining their morale. The domestically developed R18 UAV (Fig. 12) used by Ukraine in the ongoing war with Russia has already caused approximately $130 million in losses of various types of enemy material, which translates to around $670 in military assets destroyed for every dollar spent on producing the drone. The R18 drone can drop grenades from a height of 100-300 meters, effectively hovering over the target. It utilizes Soviet

cumulative anti-tank grenades RKG-3 or RKG-1600 as bombs. Equipped with a thermal imager, the R18's attacks are unpredictable and highly effective, as demonstrated in the video images, where a multiple-rotor drone tracks and attacks Russian tanks. Ukraine's adoption of drone technology highlights its adaptability and innovation in response to the evolving nature of warfare. By embracing emerging technologies, Ukraine has demonstrated its ability to exploit new avenues for strategic advantage and potentially inspire other nations to follow suit. In contrast, there is a high demand for C-UAS solutions in the modern battlefield.

### D.  Analysis using technology points

Why does the drone radar detection systems at Gatwick Airport function poorly, and why is it also ineffective in the context of the Ukraine war? One of the primary reasons is the range problem. According to the Drone Dome system, it can be operated by a single operator to detect, track, identify, and neutralize hostile drones. The system has a detection range of 3.5 km for a target with a RCS of 0.002 m². However, the identification distance extends beyond 3 km using CCD/IR automatic video motion detection (VMD) and Automatic Target Recognition (ATR) technologies. This indicates that the Drone Dome system utilizes radar for detecting potential drone signals and employs EO/IR for identification purposes.

Firstly, the recognition range of 3 km represents the effective alert range of the C-UAS solution. However, it is insufficient to cover a significant airspace area. As depicted in Fig. 11, the 3 km recognition range may be longer than the runway but considerably shorter than both the Runway Protection Zone (RPZ) and the outer circle in Fig. 11.

Secondly, there is a range latency between the short detection range and the recognition range. This latency results in a time delay when transitioning from the radar sensor to the EO/IR sensor. Consequently, it increases the overall response time of the C-UAS solution. For instance, the Switchblade 600, a commonly used one-way attack (OWA) drone, has a final-attack speed of 185 km/h. Extending the detection and identification range by 6 km would provide an advanced alert time of 120 seconds, while the current 3 km range offers only 60 seconds. In summary, a longer detection and identification range enables earlier alert times, highlighting the need for a significantly greater range than the current 3 km to meet the requirements of C-UAS applications.

The core reason behind the time problem lies in the response latency between the detected echo and the final identification of the target. The latency of the detection and recognition system, denoted as $t_{CUAS}$, can be calculated as follows:

$$t_{CUAS} = DRL_{\text{radar}} + SRL_{eo} + RRL_{eo} + t_{\text{com}} \qquad (1)$$

where, $DRL_{\text{radar}}$ =the Detection Response Latency (DRL) between an echo return, detection, and eventual display of the radar sensor, $SRL_{eo}$ = Search Response Latency (SRL) between the reception and the detection of the target for the EO/IR sensor, $RRL_{eo}$ =the Identification Response Latency (IRL) between the detection, and eventual identification of the EO/IR sensor, and $t_{\text{com}}$ =the delay time costing on communication between the radar and the EO/IR sensor. The basic angular resolution of a sensor can be determined by the equation:

$$\rho_{azi} = \frac{1.22\lambda}{D} \qquad (2)$$

where, $\rho_{azi}$= the angle resolution, $\lambda$ = the wavelength, and $D$ = the antenna size. Despite the optical wavelength of an EO/IR sensor being 1/100000th that of a radar sensor, the angular resolution of the EO/IR sensor is only 1/100000th that of radar. This indicates that the searching efficiency of the EO/IR sensor is much poorer compared to the radar sensor, resulting in potential seconds of $SRL_{eo}$.

To minimize the system latency, it is preferable to reduce either $DRL_{\text{radar}}$ or $RRL_{eo}$, or ideally, both. As illustrated in Table III, compared to a second-level DRT, the millisecond-level DRT of the radar can save more time for the EO/IR sensor and the overall C-UAS solution. This reduction in latency is particularly beneficial when detecting drones with high flying speeds, such as one-way attack (OWA) drones. For example, the popular OWA drone Switchblade 600 has a final-attack speed of 185 km/h, meaning it can travel approximately 51 meters in 1 second. If the radar's $DRL_{\text{radar}}$ is in the second-level range, the EO/IR sensor may miss the location when the radar sensor sends out the alert. However, with a millisecond-level DRT of the radar, the EO/IR sensor can timely capture the location. Furthermore, by leveraging the powerful ATR function of the radar sensor, the identification task can be partially performed by the radar sensor instead of solely relying on the EO/IR sensor. Consequently, the entire $t_{CUAS}$ can be reduced to the millisecond level, resulting in a real-time WYSIWYG (What You See Is What You Get) system or a sense-and-alert system. It should be noted that the elimination of the EO/IR system in certain applications may be limited due to policy reasons, which extend beyond the scope of this paper. However, at the very least, achieving "real-time" capabilities is essential for an effective S-UAS solution.

## V.  Future Challenges

### A.  New drone targets

The development of drone detection radar and C-UAS systems must strive towards achieving a real-time sense-and-alert capability, adapting to the evolving landscape of drone technology. In particular, drone detection radar systems need to address the challenges posed by the proliferation of one-way attack (OWA) drones (or suicide drones), as shown in Fig. 13. These drones have gained prominence since Russia's 2022 invasion of Ukraine, with both countries deploying at least eight types of OWA drones. Furthermore, numerous other nations have recently acquired or are in the process of acquiring OWA drones, leading to changes in military organization, training, and defense strategies. OWA drones, also referred to as kamikaze or suicide drones, have revolutionized modern warfare by combining the precision of drone technology with the destructive capabilities of guided munitions. These compact and highly maneuverable unmanned aerial vehicles are difficult to detect and intercept. Equipped with advanced guidance systems, high-resolution cameras, and sensors, OWA drones autonomously navigate to designated targets, ensuring accurate strikes with exceptional precision.

The market for OWA drones has experienced significant growth, with the number of new models unveiled in 2021 and 2022 matching the total of the previous five decades combined. More than 120 entities from over 30 countries are involved in the development and production of OWA drones, with the United States and Israel leading the way. However, other nations are increasingly developing their own models through acquisition programs that prioritize domestically produced products. Notably, the prevalence of vertical take-off and landing (VTOL) OWA drones, which constitute over one-fourth of all models, reflects a broader trend towards lightweight, hand-carried aircraft [37]. The unique characteristics of OWA drones, including their high-speed capabilities and VTOL flying ability, present additional challenges for drone detection radar systems. It is therefore crucial to review the current available detection LSS (low, slow, and small) strategy of C-UAS solutions in order to effectively address these evolving threats.

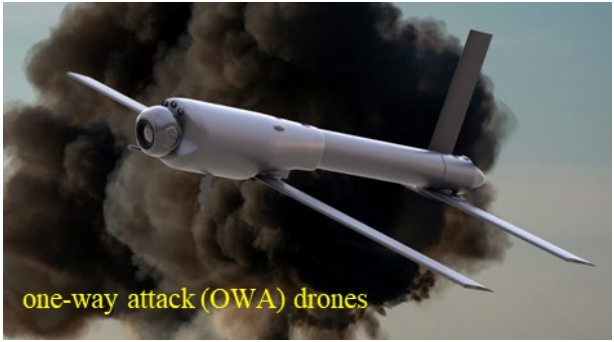

**(a)**

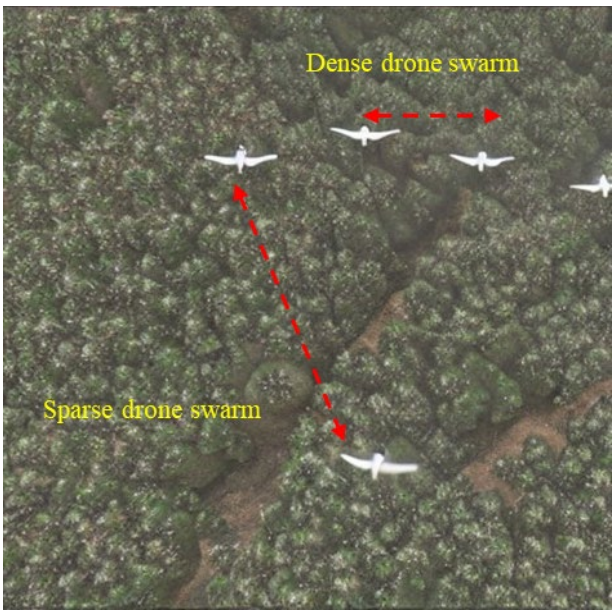

**(b)**

**Fig. 13.** The drone challenges for drone detection radar systems, (a) an example of OWA drone[1], (b) an example of sparse drone swarms[2].

1. https://www.unmannedairspace.info/counter-uas-systems-and-policies/one-way-attack-drone-market-growing-at-a-considerable-rate-new-vertical-flight-society-studyl/, accessed May, 14, 2023.

2. The image is generated using AI tools, which is provided in, https://stablediffusionweb.com/#demo.

The future of drone detection radar systems will also involve addressing the challenges posed by drone swarms. A drone swarm refers to a coordinated group of autonomous drones that work together in a synchronized manner, utilizing advanced communication and control systems [38][39]. These swarms are designed to execute complex tasks and missions that would be difficult or impossible for a single drone to accomplish alone, as shown in Fig. 13b. A key advantage of drone swarms lies in their collaborative nature and ability to share information. Each drone within the swarm acts as a node, engaging in real-time communication with other drones. This allows for the exchange of data, sharing of situational awareness, and effective coordination of actions. By employing swarm intelligence algorithms, these drones can exhibit self-organizing behaviors, collectively making decisions and adapting to changing circumstances or mission objectives.

The concept of drone swarms has garnered significant attention in both military and civilian contexts. In military applications, swarms have the potential for use in reconnaissance, surveillance, target identification, and even offensive operations. Their ability to overwhelm and confuse enemy defenses provides a substantial tactical advantage. Swarm technology also enables collaborative attacks, where multiple drones can synchronize their strikes, thereby increasing mission effectiveness and precision. Within civilian domains, drone swarms have found applications in areas such as search and rescue operations, disaster management, agriculture, and entertainment. In search and rescue scenarios, swarms can efficiently cover large areas, aiding in the detection of missing individuals or survivors. In agriculture, drone swarms can be deployed for crop monitoring, mapping, and precision spraying, optimizing farming practices. Swarms are also utilized to create captivating aerial light shows for entertainment purposes, coordinating the movement and lighting of multiple drones. Drone detection radar systems have to face the both the dense drone swarm and the sparse drone swarms. Here, the swarms only means the drones are working together in a network, but the space between the drones can be either close to each other, or far from each other. Thereby, the dense drone swarms can take part in one or two neighboring radar resolution cells, as well as intermittent radar resolution cells.

The presence of micro-Doppler signals generated by the rotating blades of drones currently plays a crucial role in the classification of radar signals. However, it raises concerns about the future scenarios where drones might lack blades altogether, resulting in the absence of micro-Doppler signals. Alternatively, the micro-Doppler signals from a drone could be weakened due to stealth blades or increased distance. In such cases, it becomes imperative to explore alternative radar signatures for the identification of radar echoes from drones. The development of effective ATR methods for drones may require consideration of additional radar signatures beyond micro-Doppler. These alternative signatures could provide valuable insights and enable accurate identification of drones even in scenarios where the micro-Doppler signals generated by blades are absent or significantly weakened. Therefore, it is crucial to investigate and leverage other radar signatures to enhance the ATR capabilities for drone detection and classification.

### B. New radar technologies

#### (1) Cognitive radar system

Cognitive radar is a novel technology designed specifically to detect small drones. It employs machine learning techniques and receiver feedback to improve detection performance, as illustrated in Fig. 14a. Coined by Simon Haykin [40], "cognitive radar" refers to the implementation of four fundamental cognitive features: the perception-action cycle, memory, attention, and intelligence [41][42][43]. The radar field has shown considerable interest in cognitive radar, leading to numerous studies exploring its capabilities. Markus Steck et al. have reported successful utilization of cognitive capabilities in the latest Hensoldt radars, including naval and ground radars, as well as sense & avoid radar prototypes by using adaptive processing to mitigate interference signals between nearby transmitted frequencies [44]. K. Barth et al. showed that employing a cognitive waveform approach yielded a 15% improvement in classification accuracy compared to a static approach using a single waveform [45]. Furthermore, A. Huizing et al. explored the potential of incorporating deep learning techniques like Convolutional Neural Networks (CNNs) and Recurrent Neural Networks (RNNs) to classify mini-drones using micro-Doppler spectrograms in the context of cognitive radar [30]. Our earlier research has indicated the feasibility of designing a "cognitive micro-Doppler radar" capable of detecting and tracking micro-Doppler signals from drones by adapting the radar dwell time [25]. In conclusion, micro-Doppler radar represents a promising approach to drone detection due to its advanced performance in radar detection, classification, and tracking, especially for small drones.

(2) Millimeter wave radar

Millimeter wave radar, with the feature of low size, weight and power (SWaP), may be the perfect ones for attaching on the military vehicles for providing drone detection. Millimeter wave radar offers a range of advantages for detecting and tracking drones, both in terms of its technological capabilities and cost-effectiveness [46][47]. Technologically, millimeter wave radar stands out due to its ability to operate in high-frequency electromagnetic waves, providing exceptional resolution and accuracy in drone detection. It can effectively identify small and fast-moving targets, making it suitable for monitoring drones in various environments, including urban areas or crowded spaces. From a cost perspective, millimeter wave radar has become increasingly affordable and accessible compared to earlier versions, thanks to advancements in technology and increased market availability. This accessibility has enabled its integration into a wider range of security systems, making it an attractive choice for drone detection and tracking. The lesson that Ukraine force attacked Russian Troops Tank & Infantry with drones indicate that modern tanks or other military vehicles require urgently shield protection from drones. The range of these micro-drones are generally not exceeding $10,000, can reach several kilometres or even tens of kilometres for the most enduring. They can also fly at very high altitudes and discreetly. Furthermore, thanks to their electric motor, these drones emit no heat and have a very low acoustic signature. Therefore, they are hardly perceptible, drastically reducing the probability of an interception. Note that, a small consumer-grid drone, such as DJI Phantom 4, with $1000 can destroy a $100 million T72 tank, which is a quite high exchange ratio. Since millimeter wave radars can be small and cost-effective, it can be attached on the modern tanks or other military vehicles. Overall, the combination of advanced technological capabilities and cost-effectiveness makes millimeter wave radar a highly desirable option for effective and efficient drone detection and tracking.

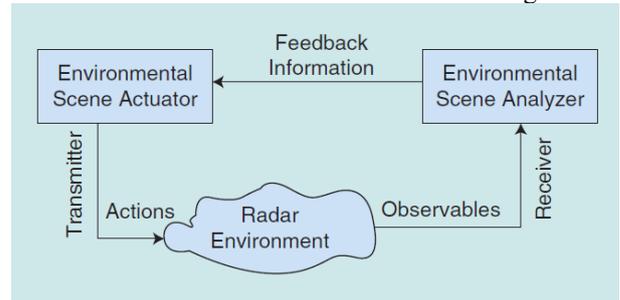

(a)

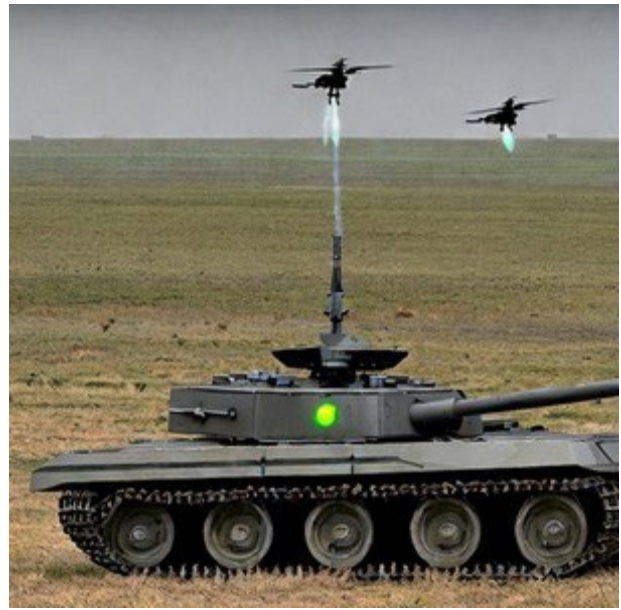

(b)

**Fig. 14.** The drone challenges for drone detection radar systems, **(a)** A block diagram of a cognitive radar seen as a dynamic closed loop feedback system with the perception–action cycle [48][42], **(b)** the Millimetre wave radar attached on a tank, which is countering drone threats[1].

1.    The image is generated using AI tools, which is provided in, https://stablediffusionweb.com/#demo.

## VI. CONCLUSION

The drone detection radar system requires ATR (Automatic Target Recognition) capabilities more than traditional air surveillance radar. This is because traditional radar operators rely on the trace and RCS to classify targets. However, drones are considered LSS (Low-Slow-Small) targets, which means they can appear as flickering ghosts and are difficult to detect, track, and classify from clutters, including birds. This poses a challenging task for human operators. Therefore, ATR for drone detection radar systems is necessary and urgent.

In the field of radar automatic target recognition (ATR), the detection and classification of small drones, particularly those belonging to Group 1 and 2, present a significant challenge. These drones fall under the recognition problem category, and the current ATR methods for small drones typically involve the analysis of kinetic features and signal signatures, such as micro-Doppler. Designing a drone detection radar system with an ATR capability requires careful consideration of adjusting radar parameters to enhance ATR signatures, guided by the scattering region theory. Creating an effective drone detection radar system necessitates adopting a comprehensive ATR approach that addresses all relevant factors and achieves a balance between the conflicting requirements of detection and tracking. This comprehensive approach ensures the maximization of the radar system's effectiveness and efficiency in accurately and reliably detecting drones. As an illustrative example, we present a drone detection radar system that demonstrates the performance improvements achieved through the integration of an ATR unit. The advantages of ATR unit bring to the drone detection radar includes: (1) enhancing beyond-range detection by providing feedback to the detection unit; (2) improving the recognition tier by recognizing sub-class of targets; (3) facilitating situation awareness by forwarding information to the tracking unit.

The increasing challenges posed by one-way attack (OWA) drones highlight the need for effective drone detection radar systems. Cognitive radar is an emerging solution to tackle these problems. With cognitive capabilities, radar systems can adapt to changing sensing tasks, optimize target detection performance, and improve situational awareness. By incorporating ATR capabilities, a 3D radar system can be elevated to a 4D radar system, providing 3D location and 1D attribution, thereby enhancing the drone detection performance. These advancements have the potential to improve radar performance across military, civilian, and commercial applications. Staying ahead in research and development enables radar systems to effectively detect, track, and respond to emerging threats, resulting in valuable insights in diverse fields.